\documentclass[a4paper,10pt]{article}
\usepackage[utf8]{inputenc}

\title{Deformation of Second and Third Quantization}
\author{Mir Faizal \\Department of Physics and Astronomy, \\  University of Waterloo,   Waterloo,\\
Ontario N2L 3G1, Canada  
}
\date{}

\begin{document}

\maketitle

\begin{abstract}
In this paper, we will deform the second and third quantized theories by deforming the canonical  
commutation relations in such a way  that they become 
consistent with the 
generalized uncertainty principle. Thus,  we will  first deform the second quantized commutator  and 
obtain a deformed version of the Wheeler-DeWitt equation. Then we will further deform the third quantized theory by 
deforming the third quantized canonical  commutation relation. This way we will obtain a deformed version of the
third quantized theory for the multiverse. 
\end{abstract}

\section{Introduction}
In the second quantized models of quantum gravity, all the 
physical information about the universe can be extracted from the wave  function of the universe as it describes the 
quantum state of the universe \cite{Hartle83}-\cite{Hawking}. 
This wave function is obtained by taking a sum over all geometries and field configurations which match with a particular field configuration 
at a spatial section of the spacetime. 
This approach is called the the Hartle-Hawking
no-boundary proposal. 
In this approach a Wick rotation to Euclidean time makes this integral well defined. 
This wave function of the universe can also be viewed as a solution of the Wheeler-DeWitt equation \cite{DeWitt67}-\cite{Wheeler57}.
The Wheeler-DeWitt equation is a second quantized equation which can be interpreted as the Schroedinger's equation for gravity. 
However, there is no time in the Wheeler-DeWitt equation because it has to satisfy the time invariance required by 
 general relativity \cite{Hawking81}.

The wave function of the universe corresponding to various other boundary conditions  has been studied. 
In the Vilenkin proposal the wave function of the universe is obtained by a quantum tunneling transition
\cite{Vilenkin86}-\cite{Vilenkin82}.  A baby universe can be created by a quantum fluctuation of the vacuum  and this baby universe 
may eventually jump into an inflationary period  and become a parent  universe. Here the parent
universes are defined to be universes with a large 
Hubble length and baby universes are defined to be virtual  fluctuations of the metric. 
In this model, there is no reason why only a single baby universe will jump into an inflationary period and  become a parent  universe. 
Thus, this model naturally leads to the existence of  multiple universes or the multiverse \cite{Linde86}. 
In fact, this model of inflationary in the multiverse is called the chaotic inflationary multiverse. 
It has been argued in this theory that the total number of distinguishable locally Friedman universes generated by
eternal inflation is proportional to the exponent of the entropy of inflationary perturbations \cite{Linde}. 
The multiverse also appears naturally in the landscape of the string
theory \cite{Davies04}. This is because there are $10^{500}$ different string theory vacuum states \cite{string}, and it is suspected that 
all these different vacuum states could be real vacuum states of different universes \cite{string0}.
 
The Wheeler-DeWitt equation is the Schroedinger's equation for gravity, and  just as the creation and annihilation of 
particles cannot be explained using the first quantized formalism, the creation and annihilation 
of universes cannot be explained using the second quantized formalism. 
So, just as we need to go to a second quantized formalism to explain the creation and annihilation of particles, we 
need to go to a third quantized formalism to explain the creation and annihilation of universes
 \cite{Strominger90}-\cite{third}. In the third quantized formalism the Wheeler-DeWitt equation is viewed as a classical field equation, 
 and a classical action corresponding to it obtained. The addition of interaction terms then 
 accounts for the creation and annihilation of universes. Thus, third quantization is a natural formalism for analysing the multiverse. 
It may be noted that the third quantization of the Kaluza-Klein theories has been studied \cite{ia}.
In doing  so  the number density of the universes created from the vacuum were calculated and an attempt was made to statistically explain the 
compactification.

It may also be noted that recently  the deformation of the first quantized theories
by a minimum measurable length has been studied. This is because string theory  predicts the existence of a  minimum  length 
 \cite{z2}-\cite{2z}. In fact, even  in loop quantum gravity  
the existence of minimum length 
 turns big bang into a big bounce \cite{z1}. 
Strong indication for the existence of a minimum length of the order of the Planck length also occur 
in  black hole physics  \cite{z4}-\cite{z5}.
However, according  to the Heisenberg uncertainty principle there
is no limit to the accuracy with which one can measure 
 the momentum or the position of a particle separately. Thus, the 
minimum observable length is  zero. If we want to incorporate
the idea of minimum length, then the Heisenberg uncertainty principle has to be modified. This, modified 
 uncertainty principle is called the 
generalized uncertainty principle 
\cite{1}-\cite{54}. In fact, we have to even modify the Heisenberg algebra to make it consistent with this 
modified  uncertainty principle. This modification of the Heisenberg algebra deforms 
all first quantized  Hamiltonians  as
$
 H \psi = H_0\psi + H_1\psi
$, where $H_0  $ is the original Hamiltonian, and 
$H_1 = \beta    p^4 /m$ is the term that occurs due to the existence of a  minimum length. 
The second quantization of this deformed first quantized theory has also been studied \cite{n9}-\cite{n}. 
In fact,  certain models of inflation motivated by such developments have also been analysed \cite{infla}-\cite{infla1}. 
  In these models the   
inflationary scalar density perturbations effectively reduces   to  
 a minimally coupled
massless real scalar field on a fixed curved background spacetime. These scalar perturbations are modified 
by the existence of a minimum length scale in the theory. 
Thus, in this work the  coordinate representation of the  momentum operator for scalar perturbations is modified, 
and this corresponds to the introduction of higher derivative terms in the action for the  minimally coupled
massless real scalar field. It is also possible to deform the second quantized commutation relation between any field theory, 
and this is expected to change the quantum mechanical aspects of such a theory,
while leaving the classical equations of motion of such a theory un-deformed . Thus, this deformation is expected 
to change the kinitic part of the  Wheeler-DeWitt equation. In fact, such a 
  deformation of Wheeler-DeWitt equation  has already been studied \cite{wheel}. 
  In this paper, we will analyse such a deformation  of the second quantized commutator by the generalized uncertainty principle. We will also 
deforming the third quantized commutator by the generalized uncertainty principle.

\section{Deformed Second Quantization}
Various implication of deformed Heisenberg algebra for quantum field theories have been studied \cite{1}-\cite{54}. 
In this section we will analyse the implication of this deformation for the Wheeler-DeWitt equation. 
The generalized uncertainty principle is consistent with the following deformed   Heisenberg algebra \cite{1}-\cite{54}
\begin{equation}
[x^i, p_j] =  i\delta^i_j +  i f^i_j (p ),
\end{equation}
where $f^i_j (p) $ is a tensor  constructed out of momentum operators. One of the most used representations of 
$f^i_j (p) $ is given by 
\begin{equation}
 [x^i, p_j] = i \delta^i_j [ 1 +  \beta p^k p_k ] + 2 i\beta p^i p_j, 
\end{equation}
where $\beta$ is a constant. 
This deformation corresponds to taking the following representation of the momentum operator 
\begin{equation}
 p_i = -i (1 + \beta \partial^j \partial_j )\partial_i. 
\end{equation}
Now we can deform the canonical commutation relation in  quantum field theory in a similar way. 
Thus, we can write the commutator of a scalar field theory as 
\begin{equation}
 [\phi (x ), \pi (y)] = i \delta( x- y )  +   i f(x- y). 
\end{equation}
where $f(x, y)$ are constructed from  momentum density conjugate to $\psi$. Now again motivated from the 
fact that one of the most common forms of the tensor function, $ f^i_j =  \delta^i_j p^k p_k + 2 p^i p_j  $, 
we write the deformed canonical commutation relation in  quantum field theory as 
\begin{equation}
 [\phi (x ), \pi (y)] = i \delta( x- y ) \left[ 1 + \beta \int dz  \pi^2 (z)  \right] + 2 i \beta \pi (x) \pi (y). 
\end{equation}
This corresponds to taking the following representation for $\pi (x) $
\begin{equation}
 \pi (x) = -i \left( 1 +  \beta \int dy dz \frac{\delta}{\delta \phi (y)} \frac{\delta}{\delta \phi (z)} \delta( z-y) \right) 
 \frac{\delta}{\delta \phi (x)}. 
\end{equation}
Thus, we observe that the deformation of the second quantized canonical 
commutation relation induces non-locality in the quantum field theory. 

Now in the Arnowitt-Deser-Misner $3+1$ decomposition of general relativity, we take the following 
 line element 
\begin{equation}
ds^{2}=g_{\mu\nu}\left(  x\right)  dx^{\mu}dx^{\nu}=\left(  -N^{2}+N_{i}%
N^{i}\right)  dt^{2}+2N_{j}dtdx^{j}+h_{ij}dx^{i}dx^{j}.
\end{equation}
where $N_{i}$ the shift function and $N$ is the lapse function.
Now we can write the  Lagrangian for a universe filled with a cosmological constant $\Lambda$ in terms   
 of the  the second fundamental form $K_{ij}$, 
  the three dimensional scalar curvature $^{3}R$ and  
the three dimensional determinant of the metric $\sqrt{^{3}h}$ as, 
\begin{equation}
\mathcal{L}\left[  N,N_{i},h_{ij}\right]  =\sqrt{-g}R=\frac{N{}\,\sqrt{^{3}h}}{2\kappa}
\left[  K_{ij}K^{ij}-K^{2}+\,\left(  ^{3}R-2\Lambda\right)  \right]  ,
\end{equation}
where   $K=$ $h^{ij}K_{ij}$ is the
trace of the second fundamental form. The conjugate momentum is defined to be 
\begin{equation}
\pi^{ij}=\frac{\delta\mathcal{L}}{\delta\left(  \partial_{t}g_{ij}\right)}
\end{equation}
The Hamiltonian can  be calculated using a Legendre transformation, 
\begin{equation}
\tilde H=
dx\left[  N {H} + N_{i}{H}^{i}\right]  ,
\end{equation}
where
\begin{eqnarray}
{H}&=& \left(  2\kappa\right)  G_{ijkl}\pi^{ij}\pi^{kl}-\frac{\sqrt
{^{3}h}}{2\kappa}\left(  ^{3}R-2\Lambda\right), 
\nonumber \\
{H}^{i}&=& -2\nabla_{j}\pi^{ji}.
\end{eqnarray}
Here $G_{ijkl}$ is 
defined by 
\begin{equation}
G_{ijkl}=\frac{1}{2\sqrt{h}}(h_{ik}h_{jl}+h_{il}h_{jk}-h_{ij}h_{kl}).
\end{equation}

Now the  two classical constraints $\mathcal{H}=0$,  and $\mathcal{H}^{i}=0$, are obtained through the equation of motion. 
The wave function of the universe $\psi [h]$, is obtained by promoting the promoting the phase space variable to operators. 
However,  we  will now apply a deformed canonical commutation relation, 
\begin{eqnarray}
 [h_{ij} (x), \pi^{kl} (y)] &=&  i \delta (x-y)
 (\delta_i^k\delta_j^l + \delta_i^l \delta_j^k)
 \nonumber \\ && \times \left[ 1 + \beta \int dz G_{nmpq} \pi^{mn}(z) \pi^{pq}(z) \right] 
 \nonumber \\ && + 2 i \beta \pi_{ij}(x) \pi^{kl} (y).  
\end{eqnarray}
It is possible to choose a more general structure for this deformed canonical commutation relation, such that 
it is consistent with the tensor index symmetry of the equation. However, we will chose this particular form 
for the  deformed canonical commutation relation to simplify further calculations. 
The momentum operator corresponding to this deformed canonical commutation relation is given by 
\begin{equation}
 \pi^{ij} (x) = -i \left( 1 + 2 \beta \int dy dz
 G_{klmn}\frac{\delta}{\delta h_{kl} (y)} \frac{\delta}{\delta h_{mn} (z)} \delta( z-y) \right) 
 \frac{\delta}{\delta h_{ij} (x)}. 
\end{equation}
Thus, we can write the deformed  Wheeler-DeWitt equation  as 
\begin{equation}
  \mathcal{H}\psi\left[ h\right]  = 0
\end{equation}
where 
\begin{eqnarray}
  \mathcal{H} &=& - \left(  2\kappa\right)  
  G_{ijkl}\left( 1 + 2 \beta \int dy dz
 G_{nmop}\frac{\delta}{\delta h_{mn} (y)} \frac{\delta}{\delta h_{op} (z)} \delta( z-y) \right) 
 \frac{\delta}{\delta h_{ij} (x)}\nonumber \\&&\times 
  \left( 1 + 2 \beta \int dy dz
 G_{qrst}\frac{\delta}{\delta h_{qr} (y)} \frac{\delta}{\delta h_{st} (z)} \delta( z-y) \right) 
 \frac{\delta}{\delta h_{kl} (x)}\nonumber \\&&
-\frac{\sqrt
{^{3}h}}{2\kappa}\left(  ^{3}R-2\Lambda\right). 
\end{eqnarray}

\section{Third Quantization}

In this section we will analyse a deformation of third quantized canonical commutation relation. 
In order to analyse the deformation of the third quantization, we need to write the theory in minisuperspace approximation. 
The   
Friedman-Robertson-Walker metric for $k=1$ is given by 
\begin{equation}
ds^{2}=-N^{2}dt^{2}+a^{2}\left(  t\right)  d\Omega_{3}^{2},
\end{equation} 
where $d\Omega_{3}^{2}$ is the usual line element on the three sphere. 
If the universe is filled with a scalar  matter fields $\phi$ and a cosmological constant $\Lambda$, then the Hamiltonian 
constraint is given by 
\begin{equation}
H = - \frac{\pi_a^2 }{a} + \frac{\pi_\phi^2 }{a^3}  + V(\phi, a), 
\end{equation}
where  $\pi_a$ is the momentum conjugate to $a$, $\pi_\phi$ is the momentum conjugate to $\phi$, and $V(\phi, a) = -a + m^2 a^3 \phi^2 + \Lambda^3 /3$. 
Thus, after  applying the deformed canonical commutation relation, the deformed momentum conjugate to $a$ and $\phi$ are given by
\begin{eqnarray}
 \pi_a &=& -i \left( 1 + \beta \frac{\partial^2 }{\partial^2 a} + \beta\frac{\partial^2 }{\partial^2 \phi}\right) \frac{\partial}{\partial a}, 
 \nonumber \\ 
 \pi_\phi &=& -i \left( 1 + \beta \frac{\partial^2 }{\partial^2 a} + \beta \frac{\partial^2 }{\partial^2 \phi}\right) \frac{\partial}{\partial \phi}. 
\end{eqnarray}
Now we can write the Wheeler-DeWitt equation as $\mathcal{H} \psi [a, \phi] =0$, where 
\begin{eqnarray}
 \mathcal{H} &=& \frac{1}{a}\left(  \frac{\partial^2}{\partial^2 a} + 2\beta \frac{\partial^4 }{\partial^4 a} + 2\beta\frac{\partial^2 }{\partial^2 \phi}
  \frac{\partial^2}{\partial^2 a} \right)
  + V(\phi, a)
 \nonumber \\ && 
 -  \frac{1}{a^3}\left( \frac{\partial^2}{\partial^2  \phi} + 2\beta \frac{\partial^2 }{\partial^2 a} \frac{\partial^2}{\partial^2  \phi} + 
 2\beta \frac{\partial^4 }{\partial^4 \phi}\right). 
\end{eqnarray}
Thus, we  obtained the  deformed Wheeler-DeWitt equation in minisuperspace approximation 
by again deforming the second quantized canonical commutation relation. 

We interpret  this deformed Wheeler-DeWitt equation in minisuperspace approximation 
as a classical field equation of a classical 
field $\psi[\phi, a]$ and view  the potential as a spacetime dependent mass term 
$V(\phi, a ) = M^2 (\phi, a )$. The Lagrangian to be third quantization  can now be written as follows, 
\begin{eqnarray}
\mathcal{L}_\psi &=&  \frac{1}{2}  \psi [\phi, a ]\left[ \frac{1}{a}\left(  \frac{\partial^2}{\partial^2 a} + 2\beta \frac{\partial^4 }{\partial^4 a} + 2\beta\frac{\partial^2 }{\partial^2 \phi}
  \frac{\partial^2}{\partial^2 a} \right)
  + M^2 (\phi, a) \right. 
 \nonumber \\ && \left. 
 -  \frac{1}{a^3}\left( \frac{\partial^2}{\partial^2  \phi} + 2\beta \frac{\partial^2 }{\partial^2 a} \frac{\partial^2}{\partial^2  \phi} + 
 2\beta \frac{\partial^4 }{\partial^4 \phi}\right)\right] \psi[\phi, a ].
\end{eqnarray}
The variation of this Lagrangian  will reproduce the deformed Wheeler-DeWitt equation. 
Now we can obtain a  momentum conjugate to $\psi$ as 
\begin{equation}
 P_\psi = \frac{\partial \mathcal{L}}{\partial \partial_a \psi }, 
\end{equation}
and so we can write the Hamiltonian as
\begin{equation}
 H_\psi = P_\psi  \frac{\partial \psi}{\partial a } - \mathcal{L}_\psi. 
\end{equation}
Thus, we have obtained Hamiltonian to be third quantized. Now we  impose the deformed third quantized canonical commutation relation 
\begin{equation}
 [\psi (\phi_1), P_\psi  (\phi_2) ]  = i \delta(\phi_1 - \phi_2) \left[ 1 + \beta \int d \phi_3 P_\psi^2   (\phi_3) \right] + 2i \beta P_\psi (\phi_1) 
 P_\psi (\phi_2). 
\end{equation}
This corresponds to taking the following representation for $P_\psi$ in the Hamiltonian $ H_\psi $, 
\begin{equation}
 P_\psi(\phi) = -i \left( 1 + \beta \int d\phi_1 d\phi_2 \frac{\delta}{\delta \psi (\phi_1)} 
 \frac{\delta}{\delta \psi (\phi_2)} \delta( \phi_1-\phi_2) \right) 
 \frac{\delta}{\delta \psi (\phi)}. 
\end{equation}
Thus, we can write finally, the deformed third quantized theory as follows, 
\begin{equation}
 i \frac{\partial \Psi }{\partial a} = H_\psi \Psi, 
\end{equation}
where $\Psi $ is the wave function of the multiverse. In absence of the matter field $\phi$, this equation becomes, 
\begin{equation}
 \left[  - \frac{1}{2} \left( \frac{\partial^2 }{\partial \psi^2 }  + 2 \beta  \frac{\partial^4 }{\partial \psi^4 }\right)+ 
\frac{\tilde \omega^2(a)}{2}  \right]\Psi(\psi ,a) =i\frac{\partial \Psi(\psi ,a)}{\partial a }. 
\end{equation}
This corresponds appears like a time dependent 
 equation for a deformed 
Harmonic oscillator with time dependent frequency.
We can also define a quantity analogous to energy for this case, such that $H_\psi \Psi = E \Psi$. 
Now  wave function is known that the wave function for the un-deformed
case can be written as \cite{s1}-\cite{s2} $|\Psi_{n_1,n_2}(x,y)>$, then the deformation corresponds to the following 
by treating the 
\begin{eqnarray}
\Delta\Psi_{n_1,n_2}( \psi, a)  &=& 
\sum_{\{m_1,m_2\}\neq \{n_1, n_2)\}}  \frac{M_{m_1, m_2, n_1, n_2 }( \psi, a)  \Psi_{m_1,m_2} ( \psi, a) }{E_{n_1,n_2}-E_{m_1,m_2}}, 
\end{eqnarray}
where 
\begin{equation}
 M_{m_1, m_2, n_1, n_2 } ( \psi, a)  =  <\Psi_{m_1,m_2}( \psi, a) |2 \beta  \frac{\partial^4 }{\partial \psi^4 }
|\Psi_{n_1,n_2}( \psi, a) > .
\end{equation}
 It is known that the conventional third quantization of the Wheeler-DeWitt equation lead to the vanishing 
of the cosmological constant \cite{colm}-\cite{colm2}. However, there as the third quantized wave function also changed by this deformation, 
it is possible that by using the deformed third quantization of the Wheeler-DeWitt equation, a small but finite cosmological 
constant might be obtained. This is what would be expected from experiments, as it has been found that 
our universe is in a state of accelerated expansion \cite{super}-\cite{super5}.

 The Wheeler-DeWitt  equation describes the evolution of the wave function in the second quantized formalism. 
To deal with the creation and annihilation of universes, we used a third quantized formalism. In the 
third quantized formalism, Wheeler-DeWitt  equation was
viewed as classical field equation of a classical field. Thus, an action for this classical field was constructed such that its equation of motion 
coincides with the second quantized Wheeler-DeWitt equation. This Lagrangian was used for calculate the momentum 
conjugate to the field variables. After obtaining the momentum conjugate to the field variables, we constructed a deformed  third quantized theory by 
deforming the third quantized canonical commutation relation. It may be noted that in the third quantized theory without matter fields, 
$\psi$ acts like the space coordinate and $a$ acts like the time coordinate. The deformed third quantized equation
obtained in this case is similar to a quantum mechanical equation 
quantized using generalized uncertainty principle. 
The existence of a minimum length can be inferred from this  generalized uncertainty principle 
\cite{1}-\cite{54}. 
Now by using the general arguments used in inferring the existence of a minimum length, we can show that the 
there exists a minimum value for the wave function in this deformed third quantized theory. 
This  minimum value for the 
wave function in the deformed third quantized theory is given by  $\psi_{min} \sim \sqrt{3\beta}\neq 0$, as $\beta \neq 0$.
So, the wave function of the universe cannot  become zero in this deformed third quantized theory. 
In other words the only way for the wave function of the universe to vanish is that the third quantized deformation 
parameter is set equal to zero, and this will reduce the deformed third quantized theory to the un-deformed third quantized theory. 
Alternatively, the non-vanishing of the wave function of the universe implies the deformation of the third quantized theory. 
\section{Conclusion}
The Wheeler-DeWitt equation is the Schroedinger's equation for a single universe. 
In the paper we obtained a deformed Wheeler-DeWitt equation by deforming 
the second quantized canonical commutation relation. The Wheeler-DeWitt equation has to be third quantized 
to obtain the wave function of the multiverse. This was done by first considering the Wheeler-DeWitt equation as
a classical field equation and then writing a Lagrangian corresponding to it. This Lagrangian was used for obtaining 
a momentum conjugate to field variables. After that the theory was third quantized by imposing canonical commutation relation. 
We also deformed these third quantized canonical commutation  relation. Thus,  we  analysed 
a deformation of both the second and third quantized canonical commutation relations. It may be noted that the 
generalized uncertainty principle has been known to deform the first quantized theories. This is what motivated a similar 
deformation of the second and third quantized canonical commutation relations. 
We also argued using this logic that the wave function of the universe cannot  vanish for a deformed third quantized theory, 
without reducing this theory to the un-deformed theory. 
It will be interesting to analyse further the implications of such a deformed quantization of field theories. 

The quantum fluctuations in spacetime at Planck scale will give rise to fuzzy structure of spacetime called the 
spacetime foam \cite{Hawking90}-\cite{Hawking88}. In this model the spacetime is  populated with virtual Lorentzian and
Euclidean wormholes. The Lorentzian wormholes are solutions to the  Einstein's equations with at least two
asymptotically flat regions. They thus connecting two separate parts either of
the same universe or of two different universes. The Euclidean wormholes are
Euclidean sectors of a Friedman spacetime.  The   quantum state of these Euclidean wormholes is obtained 
from a Wick rotation of  oscillatory universes. The fluctuation of spacetime at Planck scale can also lead to the formation of virtual black holes.
Third quantization is a natural formalism for analysing different models of spacetime foam. In fact, 
virtual black holes have been studied using third quantization  \cite{1111}. It will be interesting to analyse the effect that this 
deformed quantization can have 
on different models of spacetime foam. In fact, the conventional third quantization of the Wheeler-DeWitt equation lead to the vanishing 
of the cosmological constant \cite{colm}-\cite{colm2}. It will be interesting to study what 
happens to the cosmological constant problem using this deformed third
quantization. It has already been observed in this paper that the deformation of third quantization corresponding to 
generalized uncertainty principle, will change the wave function of the universe. Hence, it is hoped that this deformation can lead to the 
existence of a small but finite cosmological constant.

\end{document}